\documentclass[aps,twocolumn,amsmath,amssymb,superscriptaddress]{revtex4-2}
\pdfoutput=1
\usepackage[colorlinks=true,linkcolor=blue,citecolor=blue,urlcolor=blue]{hyperref}

\usepackage{graphicx}
\usepackage{graphics}
\usepackage[export]{adjustbox}
\usepackage{bm}
\usepackage{amsfonts}
\usepackage{amssymb}
\usepackage{amsmath}
\usepackage{wasysym}
\usepackage{bbold}
\usepackage{stmaryrd}
\usepackage[usenames]{color}
\usepackage{colordvi}
\usepackage{units}
\usepackage{bbm}
\usepackage{booktabs}
\usepackage{array}
\usepackage{placeins}
\usepackage{epsfig}
\usepackage{changes}
\usepackage{braket}
\usepackage{tabularx}
\usepackage[normalem]{ulem}
\newcolumntype{L}[1]{>{\raggedright\arraybackslash}p{#1}} 
\newcolumntype{C}[1]{>{\centering\arraybackslash}p{#1}} 
\newcolumntype{R}[1]{>{\raggedleft\arraybackslash}p{#1}} 

\newcommand{\be}{\begin{equation}}
\newcommand{\ee}{\end{equation}}
\newcommand{\beqn}{\begin{eqnarray}}
\newcommand{\eeqn}{\end{eqnarray}}

\usepackage{orcidlink} 

\begin{document}

\title{Random quantum Ising model with three-spin couplings}
\author{Ferenc Igl{\'o}i\,\orcidlink{0000-0001-6600-7713}}
\email{igloi.ferenc@wigner.hu}
\affiliation{Wigner Research Centre for Physics, Institute for Solid State Physics and Optics, H-1525 Budapest,  Hungary}
\affiliation{Institute of Theoretical Physics, University of Szeged, H-6720 Szeged, Hungary}
\author{Yu-Cheng Lin}
\email{yc.lin@nccu.edu.tw}
\affiliation{Graduate Institute of Applied Physics, National Chengchi University,
Taipei 11605, Taiwan}
\date{\today}

\begin{abstract}
We apply a real-space block renormalization group approach to study the critical properties of the random transverse-field Ising spin chain with multispin interactions. First we recover the known properties of the traditional model with two-spin interactions by applying the renormalization approach for arbitrary size of the block. For the model with three-spin couplings
we calculate the critical point and demonstrate that the phase transition is controlled by an infinite disorder fixed point. We have determined the typical correlation-length critical exponent, which seems to be different from that of the random transverse Ising chain with nearest-neighbor couplings. Thus this model represents a new infinite disorder universality class.

\end{abstract}

\pacs{}

\maketitle

\section{Introduction}
\label{sec:introduction}
Quantum phase transitions take place at zero temperature by varying a control parameter, such as the strength of a transverse field, and signalled by a singularity in the ground state of the system\cite{Sachdev_2011}. Quantum phase transitions occur also in one space dimension and the corresponding singularity is often related to the temperature driven critical singularities in two-dimensional classical systems\cite{RevModPhys.51.659}. 
Disorder is an inevitable feature of real materials and it could have a very profound effect on the critical behaviour of the systems\cite{Cardy_1996} and the effect of disorder is particularly strong at a quantum critical point. The theoretical description of a random quantum phase transition is very challenging since the combined effect of disorder and quantum fluctuations, together with strong correlations have to be considered at the same time. For some models this type of investigations can be performed by a so called strong disorder renormalization-group (SDRG) method
\cite{IGLOI2005277,igloi_monthus2}.
In the SDRG calculation local degrees of freedom with a large excitation energy are successively eliminated and new parameters are calculated perturbatively for the remaining degrees of freedom. In a class of models the distribution of the renormalised parameters broaden without limit and the random phase transition is controlled by a so called infinite disorder fixed point (IDFP)\cite{FISHER1999222}. At an IDFP disorder fluctuations are overwhelmingly dominant over quantum fluctuations and the obtained critical properties are expected to be asymptotically exact for large systems.

The SDRG approach has been introduced by Ma, Dasgupta and Hu\cite{PhysRevLett.43.1434,PhysRevB.22.1305} and later applied by D. Fisher\cite{PhysRevLett.69.534,PhysRevB.51.6411} to study the critical properties of the random transverse-field Ising chain. Several presumably exact results have been obtained, so that the random model looks to be understood at least at the same level as its nonrandom counterpart. In one dimension the method has been generalised for other models resulting in exact solutions\cite{PhysRevB.50.3799,PhysRevLett.76.3001,PhysRevLett.79.3254,PhysRevLett.78.1783,PhysRevLett.87.277201}. In higher dimensions the method is applied numerically\cite{PhysRevB.61.1160,10.1143/PTPS.138.479,Karevski2001,PhysRevLett.99.147202,PhysRevB.77.140402,PhysRevB.80.214416} and different technical simplifications are introduced in order to treat large finite systems\cite{PhysRevB.82.054437,PhysRevB.83.174207,Kovacs_2011}. For the transverse Ising model it is demonstrated that the critical behaviour is controlled by IDFP-s even at higher dimensions and it is expected that the upper critical dimension of the problem is infinite\cite{PhysRevB.83.174207,Kovacs_2011}. Regarding models with nearest neighbour interaction and having a discrete order parameter the critical behaviour is expected to be the same as that of the transverse Ising model. This result seems to hold also for random stochastic models, such as the random contact process\cite{PhysRevLett.90.100601,PhysRevE.69.066140,PhysRevE.72.036126}. For random models with long-range interactions, however, a new type of fixed point is found to control the critical behaviour, which is of conventional disorder type\cite{Juhasz_2014,PhysRevB.93.184203}.

In this paper we consider another type of transverse Ising models, which have multiple-site product interactions. The Hamiltonian is defined as:
\be
{\cal H}^{(m)}=-\sum_iJ_i \prod_{l=0}^{m-1} \sigma_{i+l}^z-\sum_i h_i \sigma_i^x\;,
\label{H_m0}
\ee
%
in terms of the Pauli-matrices $\sigma_i^{x,z}$ at site $i$.
Here the interactions $J_i$ and the transverse fields $h_i$ are both independent random variables.
This model in pure (nonrandom) case has been introduced by Turban\cite{LTurban_1982} and
independently by Penson \textit{et al}\cite{PhysRevB.26.6334}. The special case
$m=2$ is the standard transverse Ising chain with nearest-neighbour
interaction. According to numerical studies the pure $m=3$ model has a
second-order quantum phase transition, which belongs to the universality class
of the 4-state Potts model\cite{FIgloi_1983,FIgloi_1986,BLOTE1986395,PhysRevB.34.4885,MKolb_1986,HWJBlote_1987},.
At the critical point there are logarithmic corrections, which are probably in
the same form as in the 4-state Potts
model\cite{CVanderzande_1987,FIgloi_1987}. For $m \ge 4$ the phase-transition
in the pure model is of first order.

In this paper we are going to study the phase transition in the disordered model with random $J_i>0$ and $h_i>0$. \textcolor{black}{We do not specify the form of the disorder distributions, we assume that these are not singular, so that the first and second moments of the log-variables are finite}. In this respect we have detailed information about the $m=2$ model\cite{IGLOI2005277,igloi_monthus2}, but the models with $m>2$ has not been investigated yet. Here we study the model with $m=3$ and consider random positive three-spin couplings and random positive transverse fields. Due to the different type of local interactions the application of the the standard SDRG method\cite{IGLOI2005277,igloi_monthus2} exhibits problems: by eliminating a strong three-spin coupling several new renormalized interactions will be generated depending on the neighbourhood of the eliminated coupling;
in this case one can not stop the proliferation of the renormalised parameters during the renormalization procedure. Therefore we chose a different approach and use a block renormalization-group method, which has been introduced to the pure $m=2$ Ising chain by Fernandez-Pacheco\cite{PhysRevD.19.3173}. This type of renormalization preserves the self-duality of the model and reproduce the exact critical point and the value of the exact correlation length exponent, $\nu(m=2)=1$. Later the method was used for other quantum spin chains\cite{HORN1980467,BHu_1980,PhysRevB.24.218,PhysRevB.24.230,PhysRevB.28.2785,FIgloi_1984,PhysRevB.48.58} and has been generalised for higher dimensions\cite{PhysRevB.24.310,PhysRevE.83.051103}. For the random transverse Ising model with nearest neighbour couplings ($m=2$) the method has been applied by Miyazaki and Nishimore\cite{PhysRevE.87.032154}, as well as by C\'ecile Monthus\cite{Monthus_2015}, both in one  and higher dimensions.

Our paper is organized as follows. In Sec.~\ref{sec:Duality}, we introduce the basic idea of the block renormalization approach and present the duality transformation of  the model. In Sec.\ref{sec:RGm2} the method is applied to the $m=2$ model by using arbitrary large size of the block, while in Sec. \ref{sec:RGm3} it is applied to the $m=3$ model with a block size $b=2$. In Secs.~\ref{sec:discussion} we close our paper with a discussion.


\section{Basic idea of block renormalization}
\label{sec:Duality}

In the block renormalization method\cite{doi:10.1139/p81-078,Pfeuty1982}  the spins at sites $i=nb+1$ with $n=0,1,2,\dots$ and $b=2,3,\dots$ are fixed at arbitrary positions while the intermediate spins are integrated out. Here $b$ sets the scale factor, which is the size of the blocks. In this way the Hamiltonian is divided into two parts:
\be
{\cal H}^{(m)}={\cal H}_0^{(m)} + {\cal V}^{(m)}\;,
\label{Hdivided}
\ee
where ${\cal H}_0^{(m)}$ represents blocks with the intra-block terms, and ${\cal V}^{(m)}$ is the perturbation that contains the transverse field acting on the selected spins and the couplings which couple the neighbouring blocks. The block-Hamiltonians are solved either analytically or numerically and the lowest levels are retained and identified as the states of the block-spin variable. At the same time the renormalized values of the inter-block terms are obtained in a (first-order) perturbative way.

In the use of the block renormalization approach it is useful if the models have duality properties in which case the number of parameters generally does not increase during renormalization. The Hamiltonian in Eq.(\ref{H_m}) indeed has such a symmetry. Following the method by Turban\cite{LTurban_1982} one can
define a set of variables:
\begin{align}
\mu_i^x&=\prod_{l=0}^{m-1} \sigma_{i+l}^z \nonumber \\
\mu_i^z&=\prod_{n=0}^{\infty} \sigma^x_{i-nm} \sigma^x_{i-nm-1}\;,
\label{mu}
\end{align}
which satisfy the Pauli spin algebra. In terms of these new variables the Hamiltonian is expressed as:
\be
{\cal H}^{(m)}=-\sum_ih_{i} \prod_{l=0}^{m-1} \mu_{i-l}^z-\sum_i J_i \mu_i^x\;,
\label{H_m}
\ee
which is in the same form as that in Eq.(\ref{H_m0}) with the correspondences $J_i \to h_{i+m-1}$ and $h_i \to J_i$.
It follows that the pure model with $J_i=J, h_i=h, \forall i$ is self-dual and the self-dual point $J/h=1$ corresponds to the phase-transition point.

In the following we first present the block renormalization for the $m=2$ model for arbitrary size of the block\cite{PhysRevB.48.58}, then we consider the model with $m=3$ and solve the renormalization approach for a block of two sites.

\section{Block renormalization approach of the $m=2$ model}
\label{sec:RGm2}
Here we first turn the spin variables $\sigma_i^x \leftrightarrow \sigma_i^z$ and
for convenience the sites are relabelled as $i=(j,\alpha)$ where $j=1,2,\dots$ labels the blocks and $\alpha=0,1,\dots,b-1$ labels sites in a block. For the $m=2$ model the intra-block Hamiltonian is:
\be
{\cal H}_0^{(2)}=-\sum_j \sum_{\alpha=0}^{b-2}J_{j,\alpha+1} \sigma^x_{j,\alpha} \sigma^x_{j,\alpha+1}-\sum_j \sum_{\alpha=1}^{b-1}h_{j,\alpha} \sigma^z_{j,\alpha}\;,
\ee
while the perturbation is given by:
\be
{\cal V}^{(2)}=-\sum_j J_{j-1,b-1} \sigma^x_{j-1,b-1} \sigma^x_{j,0}-\sum_j h_{j,0} \sigma^z_{j,0}\;.
\ee
The leftmost spin of a block is fixed, thus the $x$-component of this spin is $\pm 1$, which fixes the sign of the magnetisation at the other sites of the block, too. We use this sign to characterise the state of the block. Solving the ground state of the block both with $+$ and $-$ leftmost spins the interaction energy between two neighbouring blocks is given in first-order perturbation theory as:
\be
\epsilon_j=-J_{j,b-1} \langle \sigma^x_{j,b-1}\rangle \langle \sigma^x_{j+1,0} \rangle= - J^R_{j} \Sigma^x_{j}\Sigma^x_{j+1}\;,
\ee
where  $\langle \sigma^x_{j+1,0} \rangle=\pm 1$ is fixed, 
and $\langle \sigma^x_{j,b-1}\rangle=m_{j}^{(b)}$ is the expectation value of the end-spin magnetisation in the $j$-th block of length $b$. 
The block-spin variables are $\Sigma^x_{j}=\pm 1$ and $\Sigma^x_{j+1}=\pm 1$, thus the renormalised value of the coupling is:
\be
J^R_{j} =J_{j,b-1} m_{j}^{(b)}\;.
\label{JR}
\ee
The end-spin magnetisation can be calculated exactly\cite{PhysRevB.30.6783,PhysRevB.48.58,PhysRevB.57.11404}:
\be
m_{j}^{(b)}=\left[ 1 + \sum_{\alpha=1}^{b-1} \prod_{k=1}^{\alpha} \left(\frac{h_{j,b-k}}{J_{j,b-k-1}}\right)^2 \right]^{-1/2}\;,
\ee
which in the simplest case with $b=2$ is given by:
\be
m_{j}^{(2)}=\left[ 1 + \left(\frac{h_{j,1}}{J_{j,0}}\right)^2 \right]^{-1/2}\;.
\ee
The renormalized value of the transverse field is obtained through duality, which amounts to interchange couplings and fields and also the two ends of the block leading to:
\be
h_{j}^R=h_{j,0} \tilde{m}_j^{(b)}\;,
\label{Jh2}
\ee
with
\be
\tilde{m}_j^{(b)}=\left[ 1 + \sum_{\alpha=1}^{b-1} \prod_{k=1}^{\alpha} \left(\frac{J_{j,k-1}}{h_{j,k}}\right)^2 \right]^{-1/2}\;,
\label{mtilde}
\ee
and
\be
\tilde{m}_j^{(2)}=\left[ 1 + \left(\frac{J_{j,0}}{h_{j,1}}\right)^2 \right]^{-1/2}\;.
\ee
Let us consider the ratio:
\be
K_{j,\alpha}=\frac{J_{j,\alpha}}{h_{j,\alpha}}\;,
\label{ratio}
\ee
and calculate its value with the renormalized parameters which is given by:
\be
K^R_{j}=\frac{J^R_{j}}{h^R_{j}}=\frac{\prod_{\alpha=0}^{b-1}J_{j,\alpha}}{\prod_{\alpha=0}^{b-1}h_{j,\alpha}}=\prod_{\alpha=0}^{b-1} K_{j,\alpha}\;.
\label{KR}
\ee

\subsection{Pure model}

For the pure model with $K=K_{j,\alpha},\,\forall j,\alpha$, Eq.~(\ref{KR}) leads to the simple relation\cite{PhysRevB.48.58}:
\be
K^R=K^b
\ee
having the fixed point $K^*=(J/h)^*=1$, which corresponds to the self-dual point of the system. Furthermore the thermal eigenvalue of the transformation is $\lambda_t=b$ and the correlation-length critical exponent is $\nu^{\text{pure}}=\ln b/\ln \lambda_t=1$. This is the exact value and in this way it has been obtained for any value of $b$.

\subsection{Random model}

Now we turn to the disordered model with random couplings and random fields.
The renormalization equation in Eq.(\ref{KR}) has a simple addition form in terms of the log-variables:
\be
\ln K^R_{j} = \sum_{\alpha=0}^{b-1} \ln K_{j,\alpha}\;.
\label{KRlog}
\ee
Interestingly, there are two scale factors, $\tilde{b} < b$, such that $\tilde{b}^N = b$.
Repeating the iterations with the scale factor $\tilde{b}$ $N$-times  ($N=2,3,\dots$),  
the form of the renormalized $K$-parameter is the same as if the renormalization is performed with a scale factor $b$ in one step.

According to the central-limit theorem, Eq.(\ref{KRlog}) in the large-$b$ limit has the asymptotic form:
\begin{align}
\ln K^R_b &= b \left[ \overline{\ln J_{\alpha}-\ln h_{\alpha}} \right] \nonumber \\
&+ b^{1/2} \left[Var(\ln J_{\alpha})+Var(\ln h_{\alpha})\right]^{1/2} v
\end{align}
where $\overline{x}$ and $Var(x)$ denotes the mean value and the variance of the random variable $x$, respectively, and $v$ is a Gaussian random variable with mean zero and variance unity. Introducing the quantum control parameter:
\be
\delta=\frac{  \overline{\ln J_{\alpha}}- \overline{\ln h_{\alpha}} }{Var(\ln J_{\alpha})+Var(\ln h_{\alpha})}\;,
\ee
we have:
\begin{align}
\ln K^R_b &= b \delta + b^{1/2} \left[Var(\ln J_{\alpha})+Var(\ln h_{\alpha})\right]^{-1/2} v\;.
\label{lnKm2}
\end{align}
\textcolor{black}{The first term in the above equation characterizes the divergence of the typical correlation length, since by definition $b \sim \xi_\text{typ} \sim \delta^{-\nu_{typ}}$ near criticality.
From Eq.(\ref{lnKm2} )we obtain the typical correlation length exponent: }
\be
\nu_\text{typ}=1\;.
\label{nu_typ2}
\ee
At the critical point ($\delta=0$),  the fluctuations in the log-couplings grow as $b^{1/2}$, \textcolor{black}{which defines the log-excitation energy:$ \ln \epsilon$. The corresponding scaling form is: $\ln \epsilon \sim b^{\psi}$ with the critical exponent:}
\be
\psi=\frac{1}{2}\;.
\ee
\textcolor{black}{This type of dynamics is relevant at an infinite disorder fixed point.}

The combination of the first and the second terms in Eq.~(\ref{lnKm2}) with  \textcolor{black}{$b \delta \approx - b^{1/2} \left[Var(\ln J_{\alpha})+Var(\ln h_{\alpha})\right]^{-1/2}$ will result in the vanishing of the leading contribution of these terms, which is connected to a  finite-size correlation exponent~\cite{Monthus_2015}, $\nu_{\text{FS}}$ defined by $\delta \sim b^{-1/\nu_\text{FS}}$, and given by:}
\be
\nu_\text{FS}=2\;.
\ee 
These results follow directly from the RG equations in Eqs.(\ref{JR}) or (\ref{Jh2}) and from the known scaling properties of the end-spin magnetisation\cite{PhysRevB.57.11404}. We can also recover known results in the off-critical region, in the so called Griffiths phase\cite{PhysRevLett.23.17,PhysRevLett.23.383}. Using Eqs.(\ref{Jh2}) and (\ref{mtilde}) we express the inverse square of the excitation energy as:
\be
\frac{1}{\epsilon^2} \equiv S= 1 + \sum_{\alpha=1}^{b-1} \prod_{k=1}^{\alpha} \left(\frac{J_{j,k-1}}{h_{j,k}}\right)^2\;,
\ee
which in the limit $b \to \infty$ is a Kesten variable\cite{10.1007/BF02392040}, which has a tail distribution:
\be
P(S) \sim \frac{1}{S^{1+\Delta}}\;,
\ee
for $S \gg 1$ with the exponent which is the positive root of the equation\cite{PhysRevE.58.4238,PhysRevLett.86.1343,PhysRevB.65.064416}:
\be
\overline{\left( \frac{J^2}{h^2}\right)^{\Delta}}=1\;.
\ee
Here $\Delta$ is related to the dynamical exponent $z$ in the Griffiths phase via:
\be
z=\frac{1}{2 \Delta}\;,
\ee
so that length and excitation energy are related as:
\be
\epsilon \sim b^{-z}\;,
\ee
and the asymptotic distribution of the log-excitation energy is given by:
\be
P(\ln \epsilon) \sim \epsilon^{1/z}\;
\ee
for $\epsilon \ll 1$.

\section{Block renormalization approach of the $m=3$ model}
\label{sec:RGm3}
\begin{figure}[h!]
\includegraphics[width=1. \columnwidth]{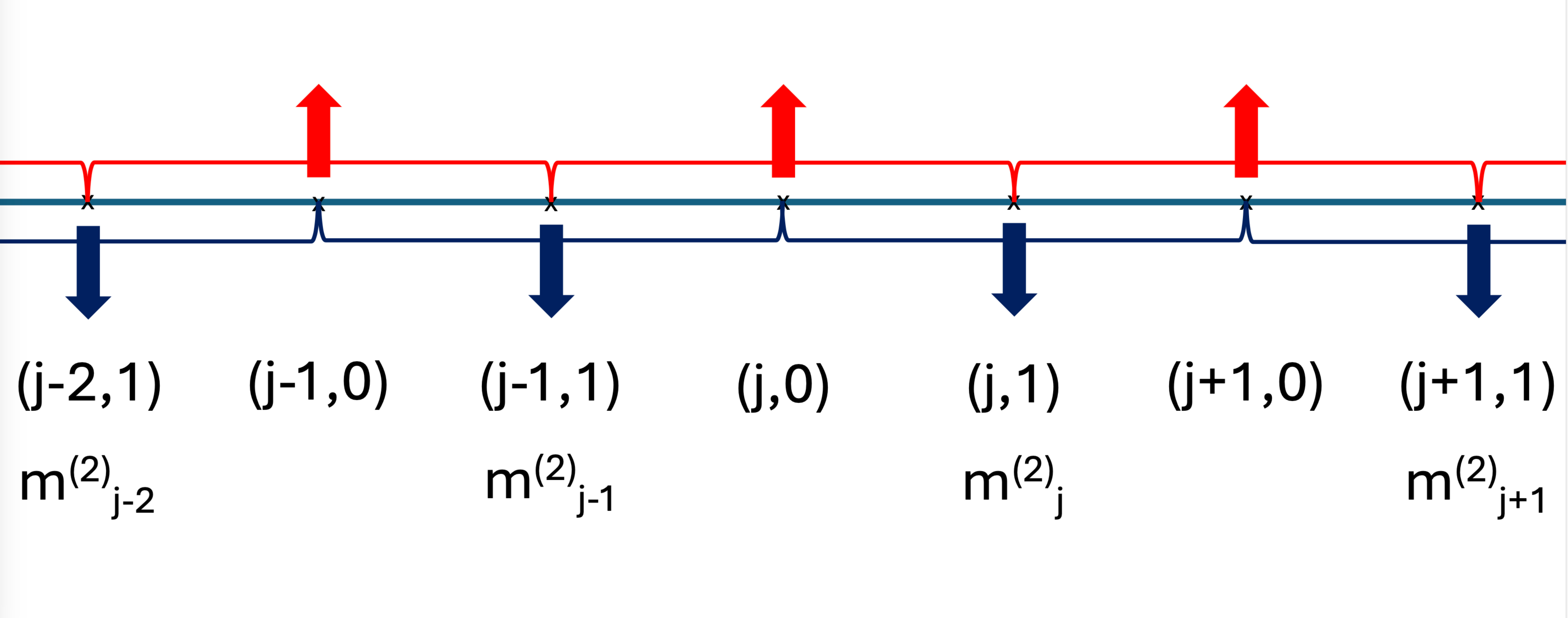}
	\vskip -0.3cm
\caption{Illustration of the division of the Hamiltonian into the unperturbed part (\textcolor{blue}{${\cal H}_0$: $\Downarrow$, $\underbrace{}$}) and the perturbation (\textcolor{red}{${\cal V}$: $\Uparrow$, $\overbrace{}$}), for transverse fields and three-spin couplings, respectively, for $m=3$ and $b=2$. The magnetization in the $j$-th block is denoted by $m^{(2)}_j$, see in Eq.(\ref{mj2}). }
\label{fig_RG}	
\end{figure}

For the $m=3$ model the Hamiltonian is split as:
\be
{\cal H}_0^{(3)}=-\sum_j \sum_{\alpha=0}^{b-2}J_{j,\alpha+1} \sigma^x_{j,\alpha} \sigma^x_{j,\alpha+1}\sigma^x_{j,\alpha+2}-\sum_j \sum_{\alpha=1}^{b-1}h_{j,\alpha} \sigma^z_{j,\alpha}\;,
\ee
and
\be
{\cal V}^{(3)}=-\sum_j J_{j,0} \sigma^x_{j-1,b-1} \sigma^x_{j,0} \sigma^x_{j,1}-\sum_j h_{j,0} \sigma^z_{j,0}\;.
\ee
Here we restrict ourselves to the case where the scale factor $b=2$, which allows the RG method to be solved analytically. 
For larger blocks $b>2$ the calculations are more complicated due to the presence of three-spin interactions.
The way how the Hamiltonian with $m=3$ is divided into two parts for $b=2$ is illustrated in Fig.\ref{fig_RG}.

Solving the ground state of the $j$th-block the interaction energy between neighbouring blocks in first-order perturbation theory is given by:
\be
\epsilon_j=-J_{j,0} \langle \sigma^x_{j-1,1}\rangle\langle \sigma^x_{j,0}\rangle \langle \sigma^x_{j,1}\rangle= -J^R_{j}  \Sigma^x_{j} \sigma^x_{j,0} \Sigma^x_{j+1}\;,
\ee
where  $\langle \sigma^x_{j,0} \rangle=\pm 1$ is fixed, $\Sigma_j^x$ denotes the block spin for the $j$th-block, and $J^R_{j}$ the renormalized three-spin coupling. 
The expectation value $\langle \sigma^x_{j,1}\rangle=m_{j}^{(2)}$ is given by:
\be
m_{j}^{(2)}=\left[ 1 + \left(\frac{h_{j,1}}{J_{j,1}}\right)^2 \right]^{-1/2}\;,
\label{mj2}
\ee
and similarly for $\langle \sigma^x_{j-1,1}\rangle=m_{j-1}^{(2)}$.
Thus the renormalized value of the three-spin coupling is given by:
\be
J^R_{j}=J_{j,0}m_{j-1}^{(2)}m_{j}^{(2)}\;.
\ee
The renormalized value of the transverse field is obtained through duality, which amounts to interchange couplings and fields leading to:
\be
h_{j}^R=h_{j,0} \tilde{m}_{j-1}^{(2)}\tilde{m}_{j}^{(2)}\;,
\label{Jh}
\ee
with
\be
\tilde{m}_j^{(2)}=\left[ 1 + \left(\frac{J_{j,1}}{h_{j,1}}\right)^2 \right]^{-1/2}\;.
\ee
Considering the ratio defined in Eq.(\ref{KR}) we obtain:
\be
K_j^{R}=\frac{J_j^R}{h_j^R}=\frac{J_{j-1,1}J_{j,0}J_{j,1}}{h_{j-1,1}h_{j,0}h_{j,1}}=K_{j-1,1}K_{j,0}K_{j,1}\;,
\ee
as a product of three original ratios. 

\subsection{Pure model}

For the pure model there is a simple relation\cite{FIgloi_1983}:
\be
K^R=K^3\;,
\ee
having the fixed-point $K^*=(J/h)^*=1$, which is the self-dual point of the system. The thermal eigenvalue is $\lambda_t=3$ and the correlation length exponent is $\nu^{pure}=\ln 2/\ln 3=0.631$.
This value is not exact but quite close to the expected value of the $4$-state Potts model\cite{Cardy_1996}: $\nu^{Potts}=2/3$.

\subsection{Random model}

For the random model repeating the renormalization in the next step we obtain:
\begin{align}
&K_j^{R(2)}=K^R_{j-1}K^R_{j}K^R_{j+1}\nonumber \\
&=K_{j-2,1}K_{j-1,0}K_{j-1,1}^2K_{j,0}K_{j,1}^2K_{j+1,0}K_{j+1,1}\;,
\end{align}
which contains the product of 9-terms, but two terms are represented twice.
Let us introduce log-ratios and iterate the renormalization process $n$-times, then we obtain an additive form:
\be
\ln K^{R(n)}=\sum_{i=1}^{L(n)} c_i^{(n)} \ln K_i\;,
\label{lnK3}
\ee
where we have used the original notation, $i$, for the sites, and $c_i^{(n)}$ is the multiplicity of the term $K_i$. For the first few iterations the multiplicities are given by:
\begin{align}
n=1\quad &111\nonumber\\
n=2\quad &1121211\nonumber\\
n=3\quad &112132313231211\nonumber\\
n=4\quad &1121323143525341435253413231211\;,
\end{align}
which can be generated using the following rules:
\be
c_{2^n}^{(n)}=1,\quad c_{2^n-i}^{(n)}=c_{2^n+i}^{(n)},\quad i=1,2,\dots 2^n-1\;
\ee
and
\begin{align}
&c_i^{(n)}=c_i^{(n-1)},\quad i=1,2,\dots,2^{n-1} \nonumber \\
&c^{(n)}_{2^{n-1}+i}=c^{(n-1)}_{2^{n-1}+i}+c^{(n-1)}_i,\quad i=1,2,\dots,2^{n-1}-1
\end{align}

Due to the scale factor $b=2$ we have for the number of sites involved in the renormalization:
\be
 L(n)=2^{n+1}-1\;,
 \ee
 and the total number of terms involving the multiplicities is:
 \be
 \sum_{i=1}^{L(n)} c_i=3^n\;.
 \label{c_sum}
 \ee
 The average value of the multiplicities is $\overline{c}=3^n/L(n)$ and its typical value has the same type of scaling with $n$: 
 \be
 c_{typ} \sim (3/2)^n\;.
 \label{c_typ}
 \ee
 Let us denote by $N_n(c)$ the number of terms having the multiplicity $c$ after $n$ iterations. According to Eq.(\ref{c_sum}) we have:
 \be
 \sum_{c=1}^{F(n)} N_n(c) c=3^n\;,
 \label{sum_c}
 \ee
where the largest multiplicity after $n$ iterations is the Fibonacci-number: $F(n)$. From Eqs.(\ref{c_typ}) and (\ref{sum_c}) follows the scaling behaviour of $N_n(c)$ as
\be
N_n(c) \sim (4/3)^n\;.
\label{scale_Nn(c)}
\ee
%

\begin{figure}[h!]
\includegraphics[width=1. \columnwidth]{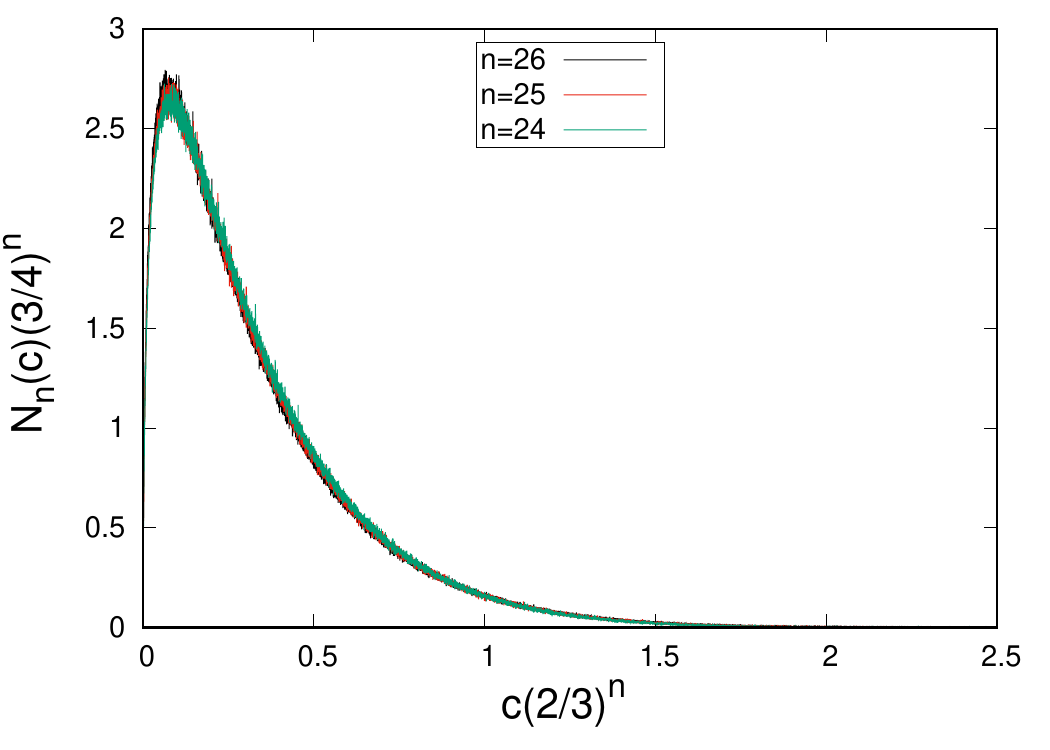}
\caption{Numerically calculated values of terms with multiplicity $c$ versus $c$ in scaled variables for different values of the number of iterations $n$. In order to reduce noise we averaged ten neighbouring values.}
\label{fig_multi}	
\end{figure} 

The numerically calculated values of $N_n(c)$ versus $c$ in scaled variables are shown in
Fig.\ref{fig_multi}. 
The points for different values
 of $n$ fit well on a master curve, confirming the scaling relations in Eqs. (\ref{c_typ}) and (\ref{scale_Nn(c)}).

Let us express the value of the renormalized ratio in Eq.(\ref{lnK3}) as:
\begin{align}
&\ln K^{R(n)}=\sum_{c=1}^{F(n)}c \sum_{k=1}^{N(c)} \ln K_k^{(c)} \approx \nonumber\\
&\sum_{c=1}^{F(n)}c \left[N(c) \overline{ln K^{(c)}}+\sqrt{N(c)\left(Var(\ln J)+Var(\ln h)\right)}v\right]\;.
\end{align}
Here  $\overline{\ln K^{(c)}}=\overline{\ln J}-\overline{\ln h} = \delta$ is the random quantum control-parameter, which is independent of the value of $c$ for the dominant part of the sum and we obtain the combination:
\be
\ln K^{R(n)} \sim 3^n \delta \sim L^{\ln 3/\ln 2} \delta\;,
\ee
 thus the typical correlation-length exponent is:
\be
\nu_\text{typ}=\frac{\ln 2}{\ln 3}\;.
\label{nu_typ3}
\ee
This value is different from that of the random transverse Ising chain with nearest-neighbour interactions, see in Eq.(\ref{nu_typ2}).

At the critical point the fluctuations of the log-couplings grow as $L^{\psi}$, with an exponent: 
\be
\psi=\frac{3\ln 3}{2\ln 2}-1=1.377\;,
\ee
 which corresponds to infinite disorder scaling.
 
 We can thus conclude that the critical behaviour of the random transverse Ising chain with three-spin couplings is controlled by an infinite disorder fixed point. The obtained value of the typical correlation length exponent indicates that the model belongs to a new infinite disorder universality class.
\section{Discussion}
\label{sec:discussion}
In this paper we have considered the transverse Ising chain with two-spin, nearest-neighbour interaction as well as with three-spin product interaction in the presence of quenched disorder and studied the quantum critical properties through a block renormalization approach. The two-spin interaction case has been well studied before and it represents the prototype system having an infinite disorder fixed point with several presumably exact results. In this case the new feature of our study that the renormalization is performed analytically for any size of the block, thus the results are correct in the large block limit. Indeed the obtained results both in the vicinity of the critical point and in the Griffiths-phase are in agreement with the previously known exact results.

On the other hand the model with three-spin couplings represents \textit{terra incognita}; to our knowledge, no random quantum systems with multi-spin interactions had been studied previously. In this case the renormalization is performed analytically for the block-size $b=2$. By iterating the transformation we noticed that the parameters at the sizes of the original model enter several times in the expression of the renormalized one. This is a new feature which is connected to the multispin topology of the interaction. We have calculated the position of the random critical point and demonstrated that the critical properties of the model are controlled by an infinite disorder fixed point. We have also determined  the typical correlation-length critical exponent, which turned out to have a different value from that of the two-spin coupling model. The latter, however represents a very broad universality class, involving  models having a discrete order-parameter and two-spin interactions.

The renormalization approach we used contains approximations, however we may expect that some of our results are asymptotically correct for an infinite disorder fixed point. Our investigations can in principle be improved by using larger blocks in the transformation, in which case, however, one should involve numerical calculations. With larger blocks, the topology of the iterated renormalization process would remain the same: the original parameters of a given site appear multiple times in the similar way, as shown for $b=2$ in this paper. Therefore we believe that the universality class of the random model with three-spin couplings is different from that having two-site interactions.

The studies presented in this paper can be extended in several directions. One could try to apply the traditional SDRG method\cite{IGLOI2005277,igloi_monthus2} by eliminating successively local degrees of freedom and explore how the topology of the renormalization equations takes a similar form, as obtained in this paper by the block renormalization approach. Another way is to perform numerical investigations of the random three-spin coupling model and check the critical properties. Performing a numerical study another critical parameters (magnetization exponent, average correlation-length exponent, etc.) and Griffiths singularities can be investigated. Studying multispin models with $m>3$ could also be interesting, since these probably represents new universality classes. One can also think to generalize the block renormalization approach to higher dimensions, a type of study that has been quite successful for the two-spin interacting model\cite{PhysRevE.87.032154,Monthus_2015}.

\textit{This work is dedicated to the memory of Ralf Kenna. Ralf was an excellent physicist and among others he studied the critical behaviour of systems above the upper critical dimension\cite{10.21468/SciPostPhysLectNotes.60}, which in our problem seems to be infinity. How the phase-transition takes place in our models at infinite dimension is still an open question.}
\bigskip
\begin{acknowledgments}
 This work was supported by the Hungarian Scientific Research Fund under Grants No. K128989 and No. K146736 and by the National Research, Development and Innovation Office of Hungary (NKFIH) within the Quantum Information National Laboratory of Hungary, and by the National Science and Technology Council (NSTC) of Taiwan under Grants No.~112-2119-M-007-008. F.I. is grateful to the Graduate Institute of Applied Physics, National Chengchi University for hospitality during his visit in Taipei.
\end{acknowledgments}

\bibliography{XYcomp}

\end{document}